\documentclass[11pt,twoside]{article}
\usepackage[utf8]{inputenc}
\usepackage{amsthm,amsmath,amssymb,epsf}

\usepackage{graphicx}

\newcommand{\dd}{{\boldsymbol d}}
\newcommand{\xx}{{\boldsymbol x}}

\begin{document}
\thispagestyle{empty}

\begin{center}
{\LARGE An integrative approach  based on probabilistic modelling and statistical inference\\
for morpho-statistical characterization of astronomical data}\\[.5in]
{\large R. S. Stoica$^{1,2}$, S. Liu$^{3}$, L. J. Liivam\"agi$^{4,6}$, E. Saar$^{4,5}$, E. Tempel$^{4}$, F. Deleflie$^{2}$, M. Fouchard$^{1,2}$, D. Hestroffer$^{2}$, I. Kovalenko$^{2}$, A. Vienne$^{1,2}$}\\[0.5in]

{\large $^1$Université de Lille, Laboratoire Paul Painlevé, France}\\
{\large $^2$Institut de Mécanique Céleste et de Calcul des Ephémérides - Observatoire de Paris, PSL Research University, CNRS,  France}\\
{\large $^3$Université Paris 1 Panthéon Sorbonne, France}\\
{\large $^4$Tartu Observatory, Estonia}\\
{\large $^5$Estonian Academy of Sciences, Estonia}\\
{\large $^6$Institute of Physics, University of Tartu, Estonia}\\
\end{center}
~\\

\begin{verse}
{\footnotesize \noindent {ABSTRACT: }
This paper describes several applications in astronomy and cosmology that are addressed using probabilistic modelling and statistical inference.\\[0.2in]


}
\end{verse}

\section{Introduction}
\noindent
Astronomical data are spatial data, in the sense that the elements of a given data set have two major components, position in an observation space and the associated characteristics, that is the measures performed at the corresponding location. Due to the data structure, it is often the case that the different problem arising induce strong morphological aspects. Therefore, the solutions to these problems can be formulated as the answer to the question "what is the pattern hidden in the data~?".\\

\noindent
Let us give a list of some examples of patterns that may occur in astronomical data: the filamentary structure outlined by the galaxy positions in our Universe, the spatial intensity of the planetary perturbations applied to comets dynamics, the confidence tube for the orbits available for a binary system or the distribution of the space debris around the Earth. The key hypothesis, that a pattern is a complex entity made of simple objects that interact, is generally accepted. Therefore, a method able to detect and to characterize these patterns should be able to extract the simple objects from the data set and to put them together in order to build the pattern. This is a typical analysis - synthesis approach. Clearly, even if the analysis part is far from being trivial, the challenge within this approach is the construction of the synthesis method, that is the integration mechanism.\\

\noindent
The common point of all the applications presented in this paper is that the chosen integrator is a probabilistic model. This choice allows not only the harmonious melting of the analysis elements into a mathematical model. But, the moments computed using the probability density of the model lead to measures that characterize the pattern, both from a statistical and morphological perspective.\\

\noindent
For the rest of the paper, let us assume that the pattern we are looking for it is the realisation of a stochastic model described by the probability density
\begin{equation}
p(\xx,\theta|\dd) = p(\xx|\dd,\theta)p(\theta),
\label{jointDensity} 
\end{equation}
with $\dd$ the observed data set, $p(\xx|\dd,\theta)$ the conditional law of the pattern and $p(\theta)$ the a priori density for the model parameters.\\

\noindent
The conditional law in~\eqref{jointDensity} can be written as
\begin{equation}
p(\xx|\dd,\theta) = \exp\left[
-\frac{U_{i}(\xx|\theta)U_{\dd}(\xx|\theta)}
{Z_{\dd}(\theta)}
\right],
\label{conditionalDensity}
\end{equation}
where the terms $U_{i}(\xx|\theta)$ and $U_{\dd}(\xx|\theta)$ are called the interaction energy and the data energy, respectively. The term $Z_{\dd}(\theta)$ determines the normalizing constant of the model probability density. In statistical physics, it is called the partition function. The interaction energy builds the pattern, while the data energy locates the pattern in the data field.\\

\noindent
Classical examples of interactions forming patterns are attraction or repulsion. For instance, two objects may attract or reject if they are to close and if they exhibit certain properties. There is a lot of freedom in defining such interactions, provided the model~\eqref{conditionalDensity} is well defined. For some specific family of models, if the interaction has a finite range, that is an object interacts only with the objects within a local neighbourhood, then the model becomes Markovian. This may lead to a simplified writing of the energy functions, known under the name of Hammersley-Clifford factorization~\cite{MollWaag04,Lies00}.\\

\noindent
The natural way of modelling using the framework described by~\eqref{conditionalDensity} is payed by the fact that $Z_{\dd}(\theta)$ cannot be always computed in an analytical closed form. The solution to this drawback is to build MCMC simulation algorithms that elude the computation of the normalising constant through $Z_{\dd}(\theta)$, hence allowing statistical inference. For instance, based on such a simulation method it is perfectly possible to build a simulated annealing algorithm, which is a global optimisation method. Under these circumstances, an estimator of the hidden pattern in the data is given by:
\begin{equation*}
(\widehat{\xx},\widehat{\theta}) = \arg\min_{\Omega \times
\Psi}\left\{ \frac{U_{\dd}(\xx|\theta)+U_{i}(\xx|\theta)}{Z_{\dd}(\theta)} -
\log p(\theta)\right\}
\end{equation*}
with $\Omega$ and $\Psi$ the corresponding state spaces for the pattern and the parameters, respectively.\\

\section{Galaxy structures distribution using marked point processes.}
\noindent
Spatial point patterns can be seen as random points configurations having random characteristics. Marked point processes are probabilistic models particularly adapted for studying such patterns~\cite{ChiuEtAl13,IlliEtAl08,Lies00,MollWaag04}.\\

\noindent
At a very large scale, the spatial distribution of the galaxies positions is not uniform~\cite{MartSaar02}. The galaxies are spread in our Universe forming amazing structures such as clusters, walls and filaments. Under the hypothesis, that the galactic filaments may be approximated by a configuration of random connected segments, the marked point processes can be used for their detection and characterisation.\\

\begin{figure}
\begin{center}
\includegraphics[width=10cm]{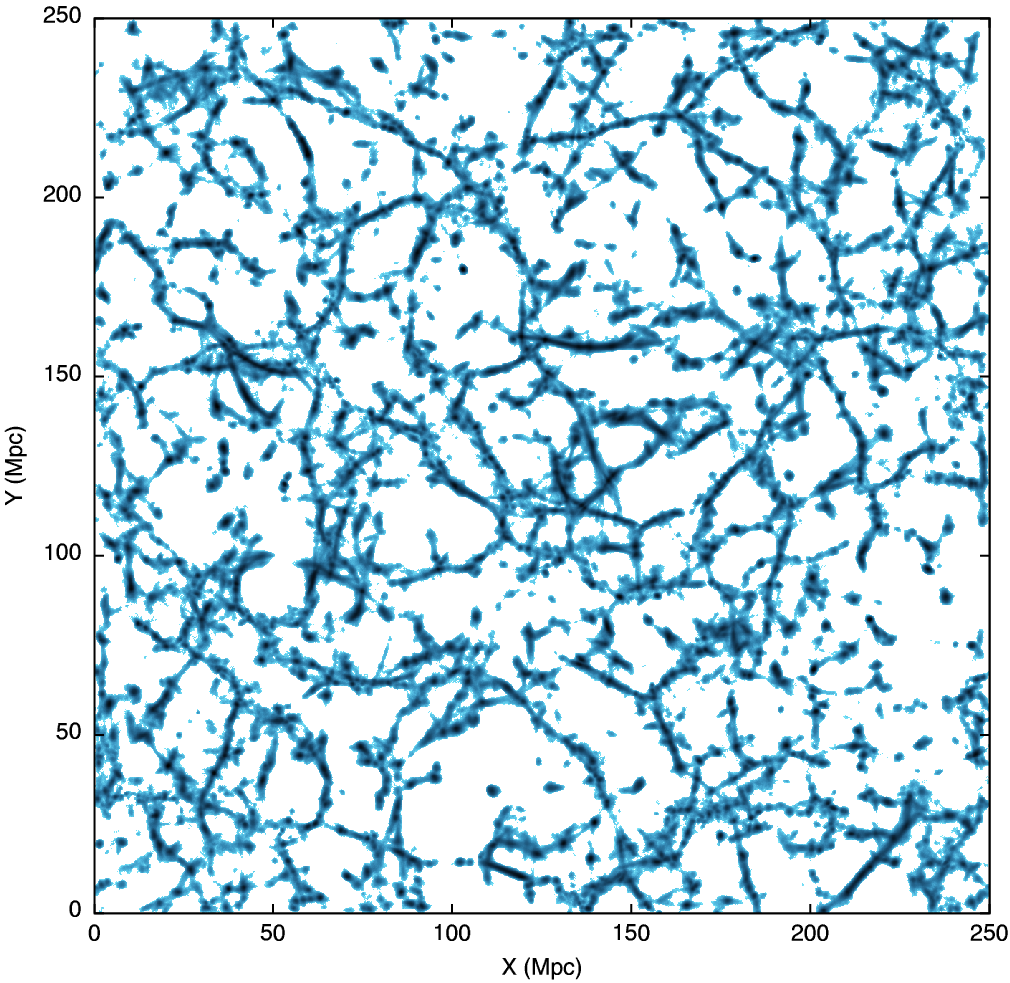}
\end{center}
\caption{Filamentary pattern detected using the Bisous model.}
\label{fig:filaments}
\end{figure}

\noindent
The Bisous model is a marked point process specially constructed for modelling general patterns made of objects that interact~\cite{StoiGregMate05}. In~\cite{StoiMartSaar07,StoiMartSaar10,TempEtAl14}, the model was successfully applied for the detection of the cosmic filaments. The Bisous model was also used for the morphological and the statistical characterisation of the detected pattern. The filaments are estimated by the segments configurations that maximise the probability density of the model. The maximisation is achieved through a simulated annealing procedure built on a tailored to the model Metropolis-Hastings dynamics. Since no particular knowledge was available for the model parameters, the chosen priors were non-informative. The morpho-statistical characterisation of the filamentary pattern was performed through the sufficient statistics of the model~: the total number of segments, the number of segments connected at one extremity and the number of segments connected at both extremities. Figure~\ref{fig:filaments} shows a detection result.\\

\noindent
Several cosmological questions were answered using the Bisous model. The authors in~\cite{TempStoiSaar13,TempLibe13} showed that the spin axes of the spiral galaxies tend to align along the galactic filaments, whereas for the elliptical galaxies, their minor semi-axes are rather orthogonal to the filaments direction. These results have direct implication for the explanation of the mechanism of galaxy formation. The connection between Bisous filaments and underlying velocity field using N-body simulations was investigated in~\cite{LibeEtAl15,TempEtAl14a}. It was found that Bisous detected filaments purely from dark matter halo and galaxy distribution are very well aligned with underlying velocity field. This confirms that the filaments detected in galaxy/halo distribution are not only visual structures, but physical systems. In~\cite{TempEtAl14b}, the galaxies distributions along the filaments were studied. For this purpose, the two point correlation function was used to show that the galaxies distribution along filaments is not uniform.\\

\section{Spatial debris distribution using spatio-temporal modelling.}
\noindent
Space debris are mostly produced by the collisions and explosions of the artificial satellites around the Earth. The aim of our undergoing project is to study the debris distribution using statistics for spatio-temporal point processes~\cite{GelfEtAl10}. The main question to answer is whether the debris behaviour exhibits a spatial pattern, such as completely random, repulsive or clustering. One of the most challenging questions tackled by this project would be the use of spatial statistics and modelling in order to find and characterize the regions around the Earth exhibiting a particular behaviour of the debris distribution. For instance, it is important to compute the probability of collision of the existing debris with a new satellite, hence obtaining an impact map for the space debris. The temporal evolution of the debris pattern is also of high interest.\\

\section{Study of the \"Opik-Oort cloud comets planetary perturbations using heavy tail distributions.}
\noindent
In the Solar System, most of the comets have very elongated orbits. Hence, the comets trajectory may cross the orbit of one of the giant planets (Jupiter, Saturn, Uranus or Neptune). If this crossing happens at a moment whenever the comet and the planet are relatively close, the comet trajectory may suffer important modifications. This mechanism is the main factor responsible of the transport of the comet in the Solar System. This mechanism is highly chaotic, so no individual motion can be easily modelled over few encounters with a planet. Our project is to consider this problem using probabilistic modelling and statistical inference.\\

\noindent
Such a study was already published in~\cite{StoiEtAl10}. The authors analysed a sample of \"Opik-Oort cloud comets affected by the four giant planets. Following the obtained results, the heavy tail distributions should be used to model the planetary perturbations around the giant planets trajectories. These perturbations tend to form a spatial pattern that is naturally explained by the \"Opik theory~\cite{Opik76}.\\

\begin{figure}
\begin{tabular}{cc}
\includegraphics[width=6.5cm]{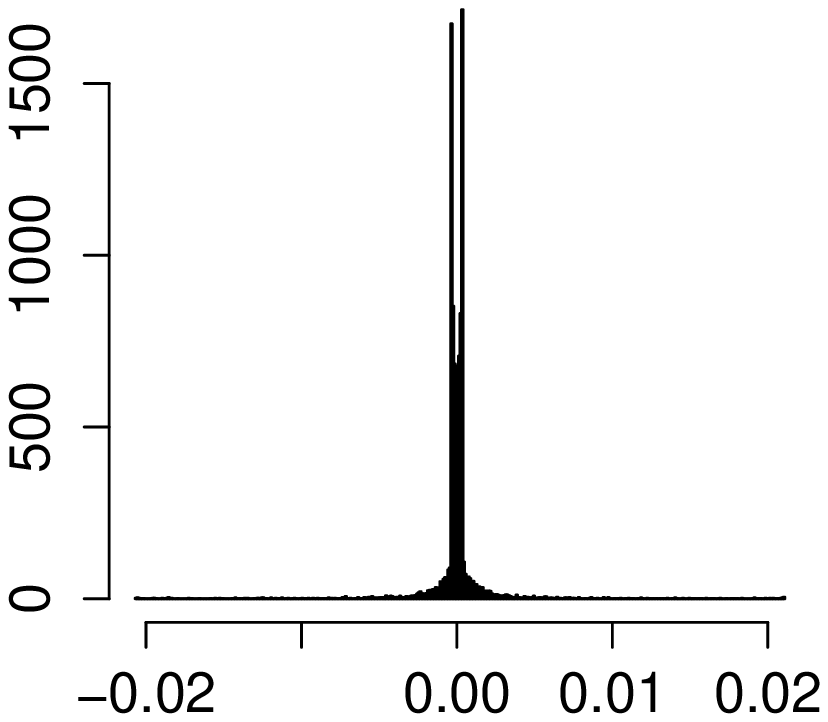} & \includegraphics[width=6.5cm]{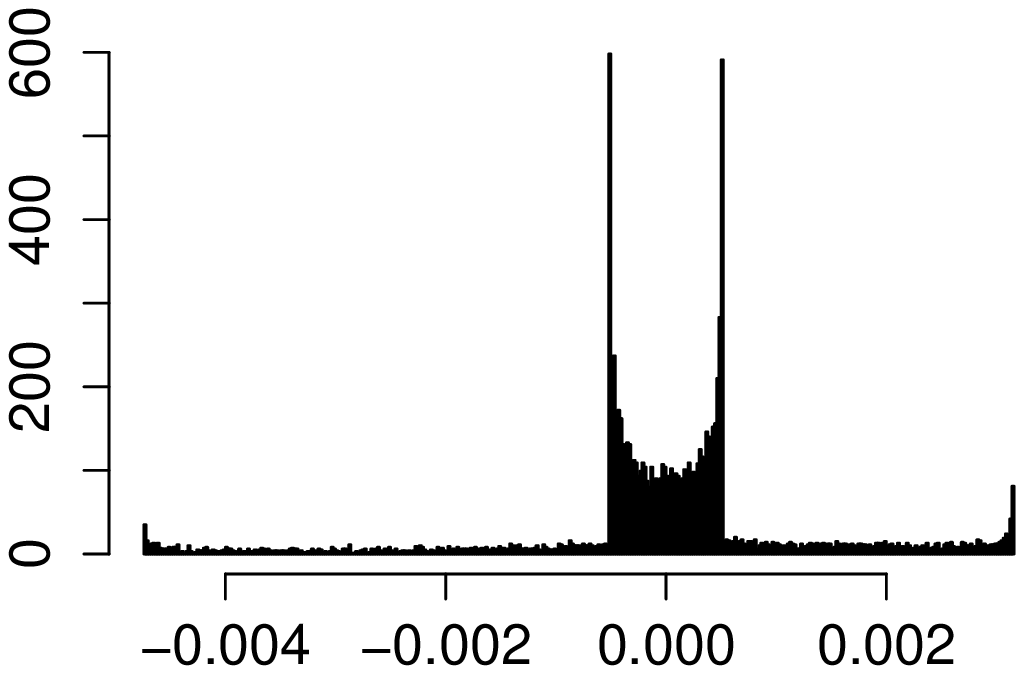}\\
\includegraphics[width=5.5cm]{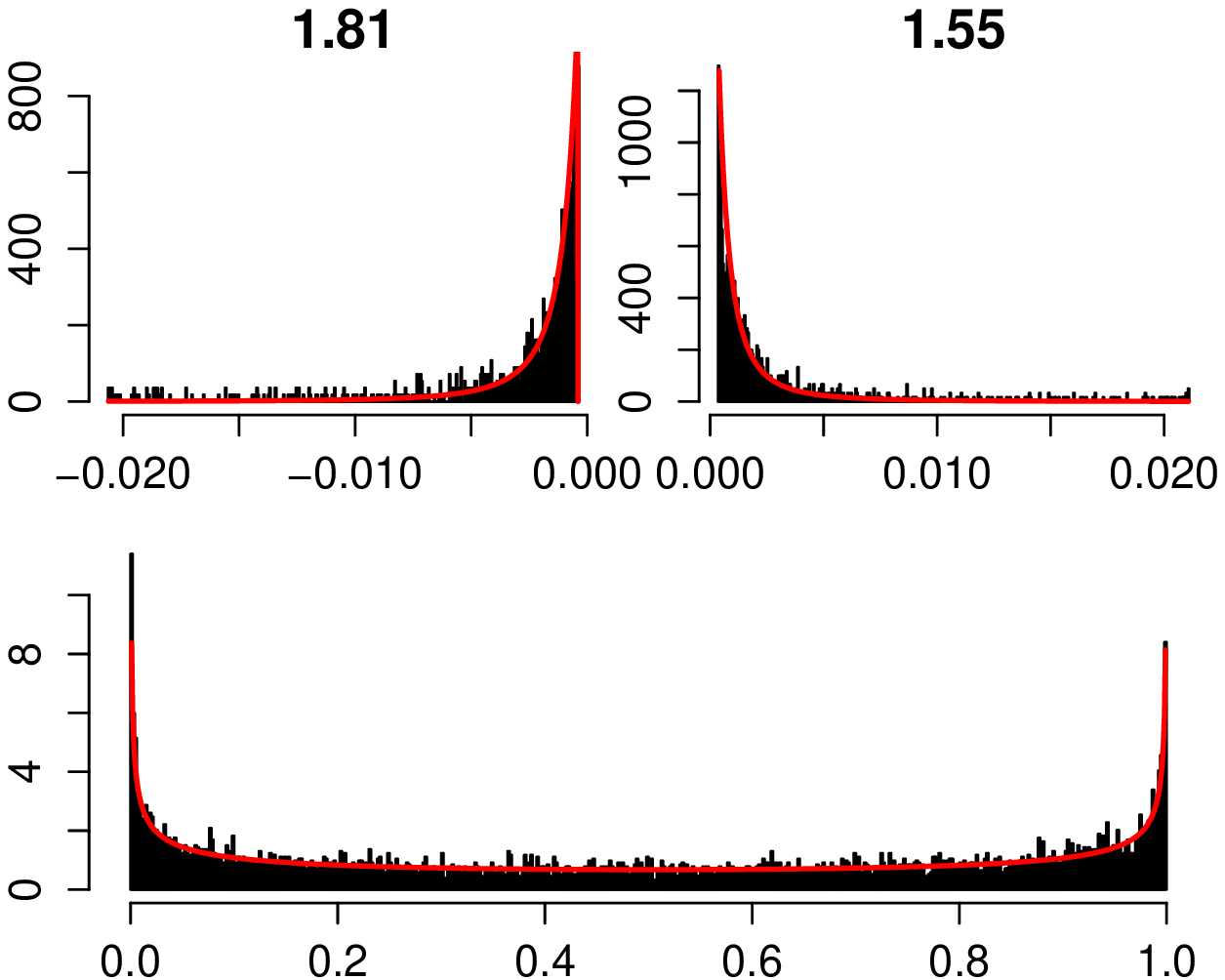}  & \includegraphics[width=5.5cm]{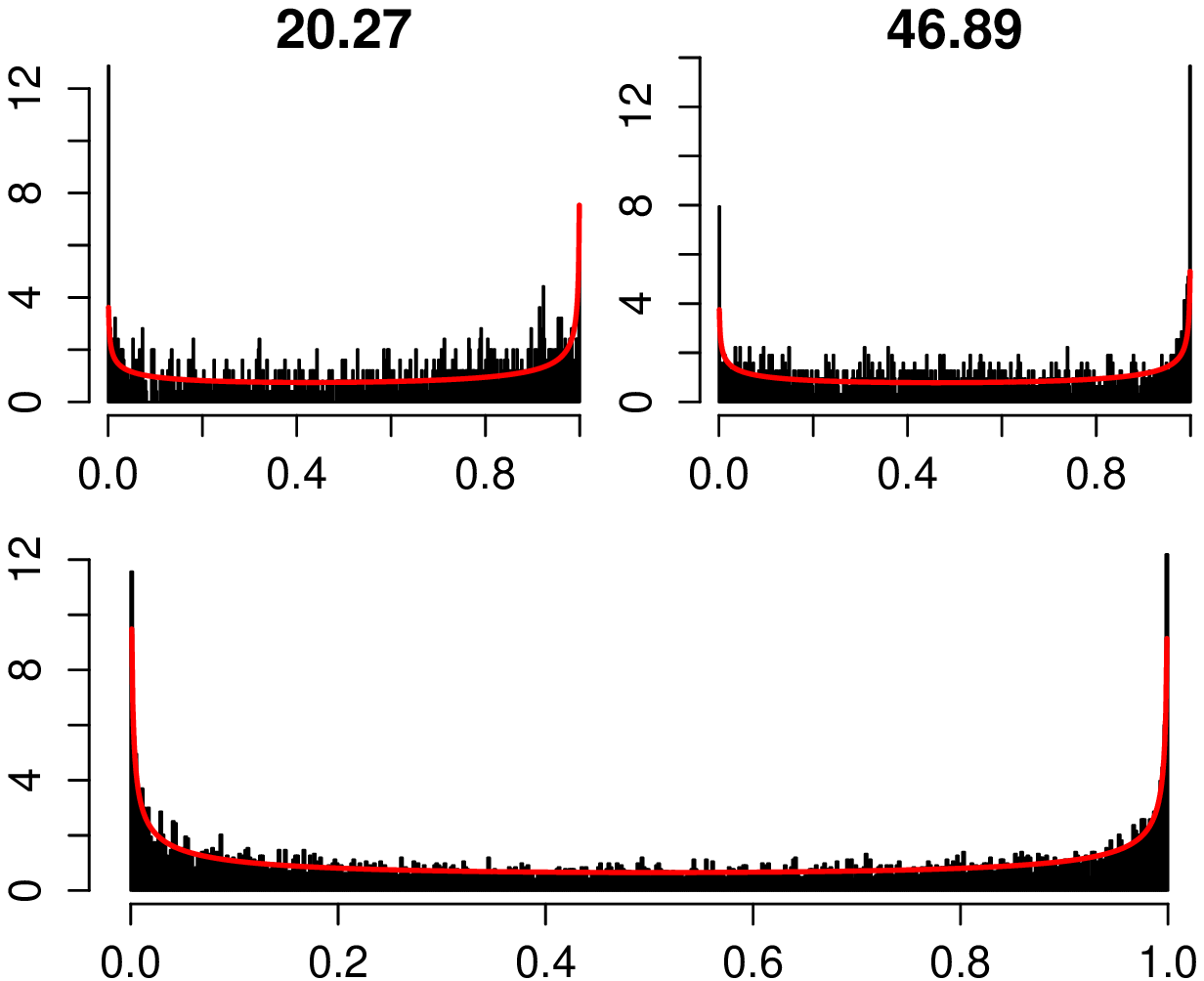} \\
\end{tabular}
\caption{Local statistical analysis for the planet perturbations. The histograms and the fitted distributions indicate that the planet perturbations distributions may poses a heavy tail character (left column) or rather a Beta distribution shape (right column).}
\label{fig:comets:analysis}
\end{figure}

\noindent
This project continues today with a study that aims to understand the separate effect of each planet on the trajectory of a comet. Therefore a simpler model composed by the Sun, a planet and the comet is under study. For the perturbations with a perihelion distance of 5.1 A.U., perihelion argument, $w$, and inclination angle, $i$, a parameter estimation was performed~\cite{StoiEtAl10}. After this, a detailed local investigation of the obtained results was done. For the perturbations that tend to exhibit a tail exponent lower than $2$, a distribution made of three components was fitted. These components are a scaled Beta distribution for the central part, and two Pareto distributions for the tails. This situation is shown in the left column of the Figure~\ref{fig:comets:analysis}. For the perturbations that tend to exhibit a tail exponent greater than $2$, again a distribution made of three components was fitted. These components were a scaled Beta distribution for the center, and Beta or Pareto distributions for the tails. This situation is presented right column Figure~\ref{fig:comets:analysis}.\\

\begin{figure}
\begin{tabular}{cc}
\includegraphics[width=6cm]{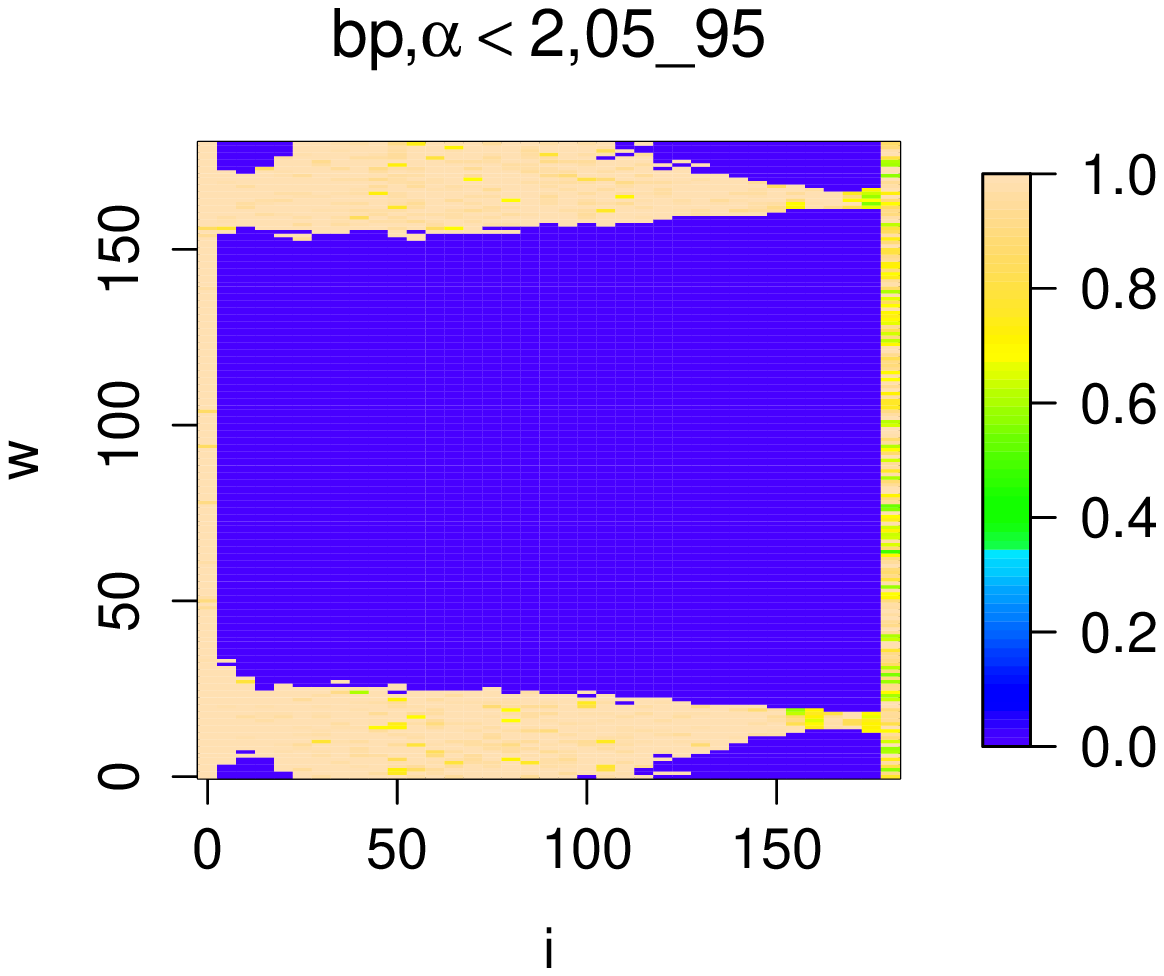} & \includegraphics[width=6cm]{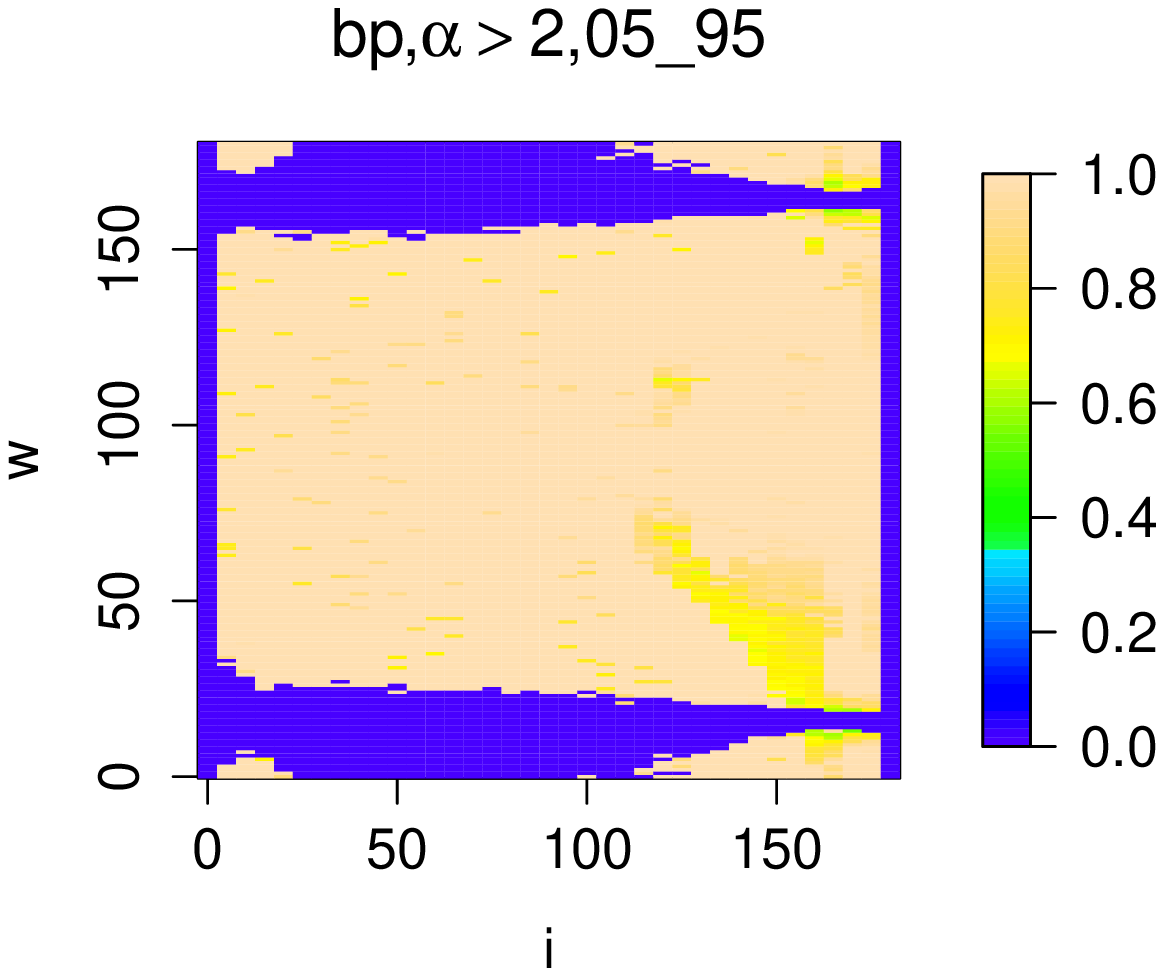} \\
\end{tabular}
\caption{Results of a simulation based statistical test, in order to validate the model fitting for the distribution of the planet perturbations. The perturbations have a perihelion distance of 5.1 A.U.. The $x$ and $y$ axes represent, the inclination angle $i$ and the perihelion argument $w$, respectively. Left: results obtained for the perturbations exhibiting a rather heavy tail behaviour. Right: results obtained for the perturbations exhibiting light or bounded tail behaviour.}
\label{fig:comets:test}
\end{figure}

\noindent
The fitted distributions were tested using a simulation based local procedure. First, the percentiles of the perturbations were estimated. Next, the perturbations were simulated $100$ times using the fitted distributions, and the percentiles were estimated each time. So, a confidence interval is obtained for each percentile. Finally, the percentage of percentiles belonging to their corresponding confidence interval is considered. The map of the obtained values is shown Figure~\ref{fig:comets:test}. The obtained results exhibit a pattern made of two "arrows" oriented from left to right. The symmetry axes of these two arrows are rather close to a perihelion argument of $w=20$ and $w=160$ degrees, respectively. These values correspond to a trajectory crossing the Jupiter's orbit. It appears, that the perturbations distributions in this region tend to exhibit a heavy tail character. In the same time, the proposed perturbation model was giving satisfactory results, for both regions, that is close and far away from the orbit of the planet. Work is still needed, in order to make all these encouraging results mathematically rigorous. The final conclusions of these two studies should be used to start the construction of a general probabilistic model for the planetary perturbations.\\

\section{Binary asteroids orbit determination through Bayesian modelling and MCMC statistical inference}
\noindent
The process of binary asteroids development and evolution started since the formation of our Solar System. Therefore, the detection and the study of such binary systems may have important consequences regarding the theoretical models of the dynamical evolution of the Solar System.\\

\noindent
Orbit determination is a classical problems of celestial mechanics. The statistical ranging method for the heliocentric orbit of a simple asteroid has been investigated  by~\cite{VirtEtAl01}. The case of the binary asteroids relative orbit determination was already tackled by~\cite{OszkEtAl09}. The idea of our approach is to use the Bayesian framework described previously in order to perform Monte Carlo statistical inference directly on the orbital parameters. This approach has the advantage that it allows the introduction of dependencies among the different orbital parameters, whenever these dependencies are a priori known. The data term of the model is built using the differences between observations and positions computed using the orbital parameters at given observation times. The a priori term is chosen to be non-informative, in the sense that for each parameter an uniform distribution is chosen. The limits of these uniform distributions are fixed for some of them, as large as possible, while for choosing the others, the observations may be used.\\ 

\begin{figure}
\begin{center}
\includegraphics[width=8.5cm]{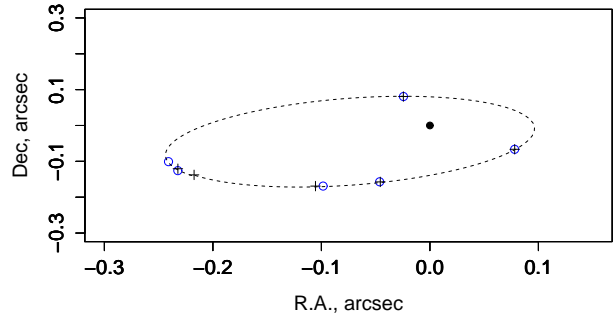}
\end{center}
\caption{Relative astrometric positions and fitted relative orbit projected onto the sky plane for the asteroid 2000QL251.}
\label{fig:orbitDetection}
\end{figure}

\noindent
Figure~\ref{fig:orbitDetection} shows a partial result of our method applied to the trans-Neptunian binary asteroid 2000QL251. The solid circle represents the primary asteroid. The open circles indicate the relative positions of the secondary asteroid. Large dotted line represents the sky plane projection of the mean fitted orbit at the mean time of the observations. The cross points are the predicted positions at the observation times given by the mean fitted orbit solution. The summary statistics given by our approach for this asteroid are shown in the Table~\ref{tab:orbitDetection}. This work is in progress. Some of perspectives we emphasize are the application of the method on data provided by the GAIA satellite. In the same, the simultaneous detection and model parameter estimation is a challenging mathematical problem.\\  

\begin{table}[]
\centering
\caption{Summary statistics for the mean orbit parameters obtained for the asteroid 2000QL251.}
\label{tab:orbitDetection}
\begin{tabular}{lllll}
                            & Min.    & Median   & Mean     & Max.     \\
Period, days                & 56.5324 & 56.5330  & 56.5330  & 56.5330  \\
Semi-major axis, km         & 4930    & 4935     & 4936     & 4943     \\
Excentrisity                & 0.4956  & 0.4958   & 0.4958   & 0.4960   \\
Inclination, deg            & 46.814  & 46.908   & 46.883   & 46.931   \\
Longitude of asc. node, deg & 75.109  & 75.125   & 75.125   & 75.206   \\
Argument of periapsis, deg  & 43.148  & 43.148   & 43.152   & 43.184   \\
Time of periapsis, RJD      & 54314.25& 54314.25 & 54314.26 & 54314.26
\end{tabular}
\end{table} 

\section{Conclusions}
\noindent
The astronomy and cosmology applications presented in this paper illustrate the extraordinary capacities of modelling and analysis of the probability and statistics based methodologies. The very strong point of these methodologies is that they come automatically with a free integrator. This integrator is the probability density of the model used to address the problem on hand. In all these situations, the model is built by the users to answer a specific question. The principle of these constructions is natural: synthesising within the models, the information gathered by a local data analysis. These models are connected with the classical mathematical models, hence allowing mutual scientific benefits for both domains, mathematics and astronomy. The common point of all the presented applications is that the solution we are looking for is represented by a pattern in a chaotic dynamics.


\bibliographystyle{plain}
\bibliography{paperHateg}

\end{document}